\documentclass[12pt]{iopart}
\usepackage{iopams}
\usepackage{graphicx}

\begin{document}

\title{Spin-orbit coupled repulsive Fermi atoms in a one-dimensional optical
lattice}
\author{Xiaofan Zhou, Kuang Zhang, Junjun Liang, Gang Chen and Suotang Jia}

\address{State Key Laboratory of Quantum Optics and Quantum Optics Devices, Institute
of Laser Spectroscopy, Shanxi University, Taiyuan 030006, China} %
\eads{\mailto{liangjj@sxu.edu.cn}, \mailto{chengang971@163.com}} \vspace{10pt%
} \begin{indented}
\item[]\today
\end{indented}

\begin{abstract}
Motivated by recent experimental development, we investigate spin-orbit
coupled repulsive Fermi atoms in a one-dimensional optical lattice. Using
the density-matrix renormalization group method, we calculate momentum
distribution function, gap, and spin-correlation function to reveal rich
ground-state properties. We find that spin-orbit coupling (SOC) can generate
unconventional momentum distribution, which depends crucially on the
filling. We call the corresponding phase with zero gap the SOC-induced
metallic phase. We also show that SOC can drive the system from the
antiferromagnetic to ferromagnetic Mott insulators with spin rotating. As a
result, a second-order quantum phase transition between the spin-rotating
ferromagnetic Mott insulator and the SOC-induced metallic phase is predicted
at the strong SOC. Here the spin rotating means that the spin orientations
of the nearest-neighbor sites are not parallel or antiparallel, i.e., they
have an intersection angle $\theta \in (0,\pi )$. Finally, we show that the
momentum $k_{\mathrm{peak}}$, at which peak of the spin-structure factor
appears, can also be affected dramatically by SOC. The analytical expression
of this momentum with respect to the SOC strength is also derived. It
suggests that the predicted spin-rotating ferromagnetic ($k_{\mathrm{peak}%
}<\pi /2$) and antiferromagnetic ($\pi /2<k_{\mathrm{peak}}<\pi $)
correlations can be detected experimentally by measuring the SOC-dependent
spin-structure factor via the time-of-flight imaging.
\end{abstract}

\pacs{03.75.Ss, 67.85.Lm}

\submitto{\NJP}

\noindent\textit{Keywords\/}: Repulsive Fermi atoms, Spin-orbit coupling,
Optical lattice


\section{Introduction}

Ultracold Fermi atoms in optical lattices have attracted considerable
interest both experimentally \cite%
{MK05,HM05,JKC06,TR06,TS06,NS07,RJ08,US08,YAL10,NS10,LT12,JSK12,DG13,JH13,GP14,PMD15}
and theoretically \cite%
{MR03,MR04,XJ05,MI05,MI130,EZ06,XG07,LM07,AM07,SJG07,FK07,BA07,MRB08,MT08,MT10,MM08,KW08,YC09,AH10,KS11,AY11,KK12,DV12,ZS12,ST14}%
, because these setups are powerful platforms to simulate rich physics of
strongly-correlated materials \cite{ML07,TS10}. One of advantages of this
system is that the spatial geometry of optical lattices can be well
controlled. Especially, using a strong harmonic transverse confinement,
one-dimensional (1D) optical lattices have been achieved experimentally \cite%
{HM05,YAL10}. On the other hand, the relative parameters have high
controllability, and moreover, can reach the regimes that cannot be
accessible in the conventional condensed-matter physics. For example, the
two-body interaction between Fermi atoms can be tuned by a
magnetic-field-dependent Feshbach resonant technique \cite{CC10}, and thus
ranges from the positive (repulsive) to the negative (attractive). For the
on-site repulsive interaction, a well-known second-order quantum phase
transition between an antiferromagnetic Mott insulator and a metallic phase
can emerge \cite{EHL68}.

Another important breakthrough in recent experiments of ultracold Fermi
atoms is to successfully create a synthetic spin-orbit coupling (SOC), with
equal Rashba and Dresselhaus strengths, by a pair of counter-propagating
Raman lasers \cite{PW12,RAW13,ZF14,LWC12}. Indeed, SOC describes interaction
between the spin and orbit degrees of freedom of a particle. In contrast to
the typical property of solid state materials that the intrinsic SOC
strength is generally smaller than the Fermi velocity of electrons, this
synthetic SOC strength realized can reach the same order as (or even larger
than) the Fermi velocity of atoms, and moreover, can also be tuned in a wide
range \cite{ZY03,KJ14}. Recent theory has revealed that SOC can generate
exotic superfluids, including topological Bardeen-Cooper-Schrieffer \cite%
{MG11,JZ11,MG12,KS12,RW12,XIJL12,MI13,HH13} and topological
Fulde-Ferrell-Larkin-Ovchinnikov \cite{CQU13,WZ13,XJL13} phases, for the
attractive Fermi atoms. The fundamental picture for achieving these
nontrivial topological superfluids is that SOC, Zeeman field, and $s$-wave
interaction can induce triplet $p$-wave pairing \cite{LPG01,CWZ08,Sato09}.
In parallel with the attractive case, it is natural to ask what novel
physics can occur in the repulsive Fermi atoms driven by the synthetic SOC
\cite{SSZ13,SSZ14,XC14}.

\begin{figure}[t]
\centering
\includegraphics[width = 3.5in]{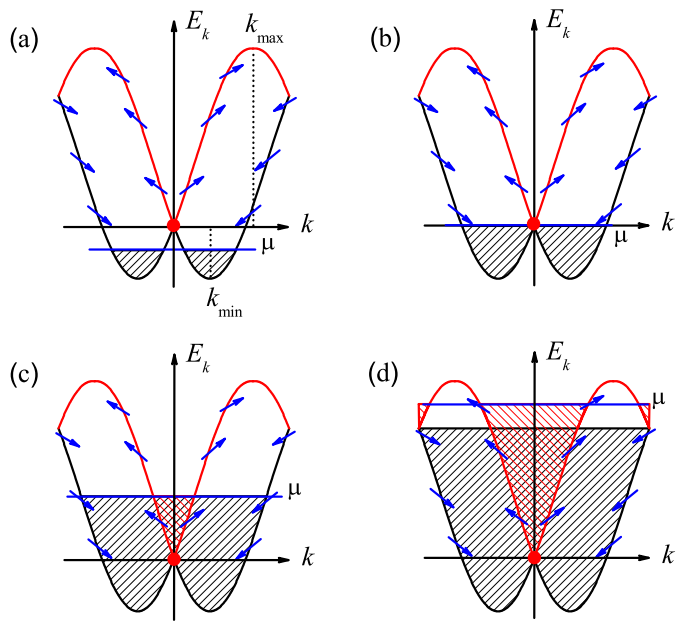}
\caption{Spin-orbit coupled energy bands in an optical lattice and the
corresponding fillings for different chemical potentials. The blue arrows
represent spin polarizations in different subbands. The red dot denotes the
contact point of two subbands. $k_{\min}$ and $k_{\max}$ are the minimum and
maximum of subbands, and are determined by Eqs.~(\protect\ref{kmin}) and~(%
\protect\ref{kmax}), respectively, if the on-site repulsive interaction is
not taken into account. (a) and (b) show the fillings for the small chemical
potentials. In these cases, only the lower subband (the black shadow) is
partly occupied. In (a), there are four Fermi points and no occupation
around $k=0$ occurs, while in (b), there are three Fermi points and the
occupation at $k=0$ emerges. (c) and (d) show the fillings for the large
chemical potentials, in which the occupations in both the lower and upper
(the red shadow) subbands emerge. In (c), the lower subband is still partly
occupied, while in (d), the lower subband is fully occupied for a larger
chemical potential. }
\label{energyband}
\end{figure}

Inspired by the above experimental developments and theoretical
considerations, here we investigate spin-orbit coupled repulsive Fermi atoms
in a 1D optical lattice. Recently, spin-orbit coupled Bose-Einstein
condensates in the 1D optical lattice has been prepared experimentally \cite%
{CH14}. Using a similar technique, the system considered could also be
achieved in the near future. Physically, the spin-orbit coupled Fermi atoms
in the optical lattices have two characteristics. One is that SOC can make
Fermi atoms hop between the nearest-neighbor sites with spin flipping; see
the Hamiltonian (\ref{HSOC}) in the following. Moreover, it has a strong
competition with the on-site repulsive interaction, and especially, with the
conventional spin-independent hopping; see the Hamiltonian (\ref{HT}) in the
following. The other is that in the presence of SOC, for different chemical
potentials, the fillings are quite different; see Fig.~\ref{energyband}.

Notice that in 1D quantum fluctuation becomes significant, and the
mean-field results are, in principle, unreliable \cite{XWG13}. Here we
capture the required ground-state properties, including momentum
distribution and spin-correlation, by using the density-matrix
renormalization group (DMRG) method \cite{LJJ14,YHC14}, which is a powerful
numerical method to study lower-dimensional strong-correlated systems \cite%
{US05}. The main results are given as follows. Section $2$ is devoted to
introducing our proposal and deriving a 1D Fermi-Hubbard model with the
synthetic SOC. Section $3$ is devoted to addressing the generalized results
without the on-site repulsive interaction. In this section, the
unconventional momentum distribution, which depends crucially on the
filling, are found. We call the corresponding phase the SOC-induced metallic
phase. Section $4$ is devoted to discussing the results in the presence of
the on-site repulsive interaction. By means of the spin-correlation
function, we find that SOC can drive the system from a spin-rotating
antiferromagnetic Mott insulator to a spin-rotating ferromagnetic Mott
insulator. As a result, a second-order quantum phase transition between the
spin-rotating ferromagnetic Mott insulator and the SOC-induced metallic
phase is predicted at the strong SOC. Here the spin rotating means that the
spin orientations of the nearest-neighbor sites are not parallel or
antiparallel, i.e., they have an intersection angle $\theta \in (0,\pi )$
(see Fig.~\ref{lattice} in the following). In the spin-rotating
antiferromagnetic Mott insulator, $\pi /2<\theta <\pi $, and the
quasi-long-range spin correlation decays as a power law and changes the sign
with a period $2<T<4$, whereas for the spin-rotating ferromagnetic Mott
insulator, $\theta <\pi /2$, and the quasi-long-range spin correlation also
decays as a power law but with the period $T>4$. Finally, we show that the
momentum $k_{\mathrm{peak}}$, at which peak of the spin-structure factor
appears, can also be affected dramatically by SOC. The analytical expression
of this momentum with respect to the SOC strength is also derived. It
suggests that the predicted spin-rotating ferromagnetic ($k_{\mathrm{peak}%
}<\pi /2$) and antiferromagnetic ($\pi /2<k_{\mathrm{peak}}<\pi $)
correlations can be detected experimentally by measuring the SOC-dependent
spin-structure factor via the time-of-flight imaging \cite{RAH15}. The
discussions and conclusions are given in section $5$.

\section{Model and Hamiltonian}

\begin{figure}[tb]
\centering\includegraphics[width = 3.5in]{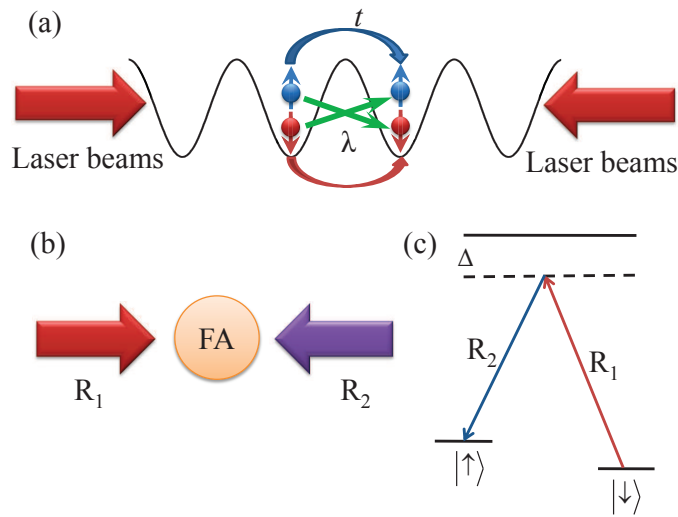}
\caption{(a) Proposed experimental setup for realizing spin-orbit coupled
repulsive Fermi atoms (FA) in a 1D optical lattice. (b) The interaction
between FA and a pair of counter-propagating Raman lasers, labeled
respectively by R$_{1}$ and R$_{2}$. These Raman lasers create a
momentum-sensitive coupling between two internal atomic states, i.e., a
synthetic SOC with equal Rashba and Dresselhaus strengths. (c) Energy levels
and their transitions induced by Raman lasers.}
\label{setup}
\end{figure}

\subsection{Proposed experimental setup}

Figure~\ref{setup} shows our proposal that repulsive Fermi atoms in a 1D
optical lattice are driven by a pair of counter-propagating Raman lasers.
For the specific experiments, 3D optical lattice is first prepared by the
interference of three pairs of counter-propagating laser beams \cite%
{ML07,TS10}. The corresponding periodic potential can be written as $V_{%
\mathrm{3D}}=V_{0}\cos ^{2}(k_{w}x)+V_{0}\cos ^{2}(k_{w}y)+V_{0}\cos
^{2}(k_{w}z)$, where $V_{0}$ is the lattice depth, $k_{w}=2\pi /\lambda _{w}$
is the wave vector, and $\lambda _{w}$ is wavelength. By further using a
strong harmonic transverse confinement $V=m_{0}\omega _{\bot }^{2}r^{2}/2$
in the 3D optical lattice, i.e., the 2D harmonic potential frequency $\omega
_{\bot }$ is far larger than the trapping frequency $\omega _{z}$ along the
weakly-confining axis, the required 1D optical lattice can be generated \cite%
{HM05,YAL10}; see Fig.~\ref{setup}(a). In such case, the two-body
interaction between Fermi atoms is described effectively by \cite{MO98}
\begin{equation}
U(z)=-\frac{2\hbar ^{2}}{m_{0}a_{\mathrm{1D}}}\delta (z),  \label{1DEI}
\end{equation}%
with the 1D $s$-wave scattering length
\begin{equation}
a_{\mathrm{1D}}=-\frac{a_{\bot }^{2}}{2a_{\mathrm{3D}}}\left( 1-C\frac{a_{%
\mathrm{3D}}}{a_{\bot }}\right) ,  \label{1DS}
\end{equation}%
where $C\simeq 1.46$, $a_{\bot }=(2\hbar /m_{0}\omega _{\bot })^{1/2}$, $a_{%
\mathrm{3D}}$ is the 3D $s$-wave scattering length, and $m_{0}$ is the
atomic mass. In addition, a pair of counter-propagating Raman lasers shown
in Fig.~\ref{setup}(b) are used to create the required 1D synthetic SOC,
with equal Rashba and Dresselhaus strengths \cite{YJL11}. In this method,
the corresponding two spin states are chosen as $\left\vert \uparrow
\right\rangle =\left\vert 9/2,-9/2\right\rangle $ and $\left\vert \downarrow
\right\rangle =\left\vert 9/2,-7/2\right\rangle $ for $^{40}$K system \cite%
{PW12}, or $\left\vert \uparrow \right\rangle =\left\vert
3/2,-3/2\right\rangle $ and $\left\vert \downarrow \right\rangle =\left\vert
3/2,-1/2\right\rangle $ for $^{6}$Li system \cite{LWC12}; see Fig.~\ref%
{setup}(c).

\subsection{Hamiltonian}

The total dynamics illustrated by Fig.~\ref{setup} is governed by the
following 1D Fermi-Hubbard model with the synthetic SOC \cite{LJJ14,YHC14}:
\begin{equation}
H=H_{\mathrm{t}}+H_{\mathrm{u}}+H_{\mathrm{soc}}  \label{TH}
\end{equation}%
with
\begin{equation}
H_{\mathrm{t}}=-t\sum_{l,\sigma =\uparrow ,\downarrow }(c_{l\sigma
}^{\dagger }c_{l+1\sigma }+\mathrm{H.c.}),  \label{HT}
\end{equation}%
\begin{equation}
H_{\mathrm{u}}=U\sum_{l}n_{l\uparrow }n_{l\downarrow },  \label{HU}
\end{equation}%
and
\begin{equation}
H_{\mathrm{soc}}=\lambda \sum_{l}(c_{l\uparrow }^{\dagger }c_{l+1\downarrow
}-c_{l\downarrow }^{\dagger }c_{l+1\uparrow }+\mathrm{H.c.}),  \label{HSOC}
\end{equation}%
where $c_{l\sigma }^{\dagger }$ and $c_{l\sigma }$ are the creation and
annihilation operators, with spin $\sigma =\uparrow ,\downarrow $, at
lattice site $l$, $n_{l\sigma }=c_{l\sigma }^{\dagger }c_{l\sigma }$ is the
number operator, $t$ is the spin-independent hopping magnitude, $U$ is the
on-site repulsive interaction strength, and $\lambda $ is the SOC strength.
Based on above proposal experimental setups, the relative parameters can be
tuned independently. For example, the hopping magnitude $t$ can be
controlled by the intensities of lasers \cite{ML07,TS10}, the 1D on-site
repulsive interaction strength $U$ can be tuned by Feshbach resonance \cite%
{CC10}, and the SOC strength $\lambda $ can be driven through a fast and
coherent modulation of Raman lasers \cite{ZY03,KJ14}. As a consequence, for
a proper optical lattice, the SOC strength $\lambda $ has the same order of
the hopping magnitude $t$ \cite{MG1205}.

\subsection{Momentum distribution and spin correlation}

In terms of the SOC-induced properties (see Introduction), here we mainly
focus on momentum distribution and spin correlation, which can be measured
experimentally by the time-of-flight imaging \cite{TR06,GP14,PW12, RAH15}.
The momentum distribution functions for spin-up and spin-down atoms are
written respectively as \cite{MO90}
\begin{equation}
n_{\uparrow }(k)=\frac{1}{L}\sum_{l,j}e^{ik(l-j)}\left\langle c_{l\uparrow
}^{\dagger }c_{j\uparrow }\right\rangle ,  \label{MDF}
\end{equation}%
\begin{equation}
n_{\downarrow }(k)=\frac{1}{L}\sum_{l,j}e^{ik(l-j)}\left\langle
c_{l\downarrow }^{\dagger }c_{j\downarrow }\right\rangle .  \label{MDF1}
\end{equation}%
For the Hamiltonian (\ref{TH}), the spin-up and spin-down atoms are equal.
It means that $n_{\uparrow }(k)$ is the same as $n_{\downarrow }(k)$, and
thus we only consider $n_{\uparrow }(k)$ in the following discussions. The
spin-correlation function is defined as \cite{RAH15,MO90,CCC08}
\begin{equation}
s(r)=\frac{1}{L}\sum_{l}\left\langle s_{l}^{z}s_{l+r}^{z}\right\rangle ,
\label{SCR}
\end{equation}%
where $r$ is a distance between different sites and $s_{l}^{z}=c_{l\uparrow
}^{\dagger }c_{l\uparrow }-c_{l\downarrow }^{\dagger }c_{l\downarrow }$. The
corresponding spin-structure factor is given by \cite{RAH15,MO90,CCC08}
\begin{equation}
S(k)=\frac{1}{L}\sum_{l,j}e^{ik(l-j)}\left\langle
s_{l}^{z}s_{j}^{z}\right\rangle .  \label{SCFM}
\end{equation}%
Since the spin-structure factor has the sum extending over all lattice sites
$l$ and $j$, it reflects spin correlation globally, and is thus used
experimentally to detect the magnetic order \cite{RAH15}.

In addition, we will perform the DMRG calculations, with open boundary
condition, to calculate Eqs.~(\ref{MDF})-(\ref{SCFM}). The basic energy
scale is chosen as $t=1$. In the detailed calculations, we retain $150$
truncated states (which is sufficient) per DMRG block and $20$ sweeps with
the maximum truncation error $\sim 10^{-5}$ \cite{MP14,SP14}.

\section{Without on-site repulsive interaction}

In order to better understand the fundamental physics induced by SOC, we
first consider a simple case without the on-site repulsive interaction ($%
U/t=0$), in which the Hamiltonian (\ref{TH}) reduces to $H_{1}=H_{\mathrm{t}%
}+H_{\mathrm{soc}}$, i.e.,
\begin{equation}
H_{1}=-t\sum_{l,\sigma =\uparrow ,\downarrow }(c_{l\sigma }^{\dagger
}c_{l+1\sigma }+\mathrm{H.c.})+\lambda \sum_{l}(c_{l\uparrow }^{\dagger
}c_{l+1\downarrow }-c_{l\downarrow }^{\dagger }c_{l+1\uparrow }+\mathrm{H.c.}%
).  \label{H1}
\end{equation}%
In experiments, the 1D noninteracting Fermi atoms can be realized by tuning
the 3D $s$-wave scattering length $a_{\mathrm{3D}}$ to its zero crossing. In
such case, the 1D $s$-wave scattering length $a_{\mathrm{1D}}\rightarrow
\infty $, and the 1D effective two-body interaction $U(z)$ then becomes zero
\cite{NS07,RJ08,TS10}; see Eqs.~(\ref{1DEI}) and~(\ref{1DS}).

\begin{figure}[t]
\centering\includegraphics[width = 4.0in]{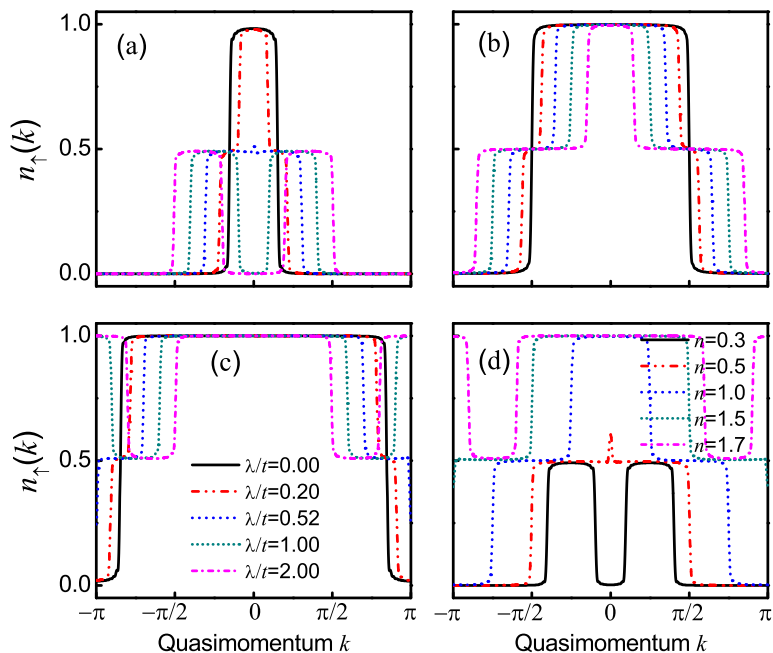}
\caption{(a)-(c) The momentum distribution functions $n_{\uparrow }(k)$ for
the different SOC strengths $\protect\lambda /t$, when the filling factors
are chosen as (a) $n=0.3$, (b) $n=1.0$, and (c) $n=1.7$. (d) The momentum
distribution functions $n_{\uparrow }(k)$ for the different filling factors $%
n$, when the SOC strength is chosen as $\protect\lambda /t=1 $. In all
subfigures, the on-site repulsive interaction strength and the lattice
length are given by $U=0$ and $L=100$, respectively.}
\label{n_k}
\end{figure}

\subsection{Momentum distribution function}

In the absence of SOC ($\lambda /t=0$), the system have three known
features. Firstly, the energy bands of the Hamiltonian $H_{\mathrm{t}}$ are
degenerate, and the system is located at the metallic phase. Secondly, all
Fermi atoms occupy the degenerate energy bands from $k_{\mathrm{min}}=0$,
where $k_{\mathrm{min}}$ is the minimum of the energy bands. Moreover, there
are two degenerate Fermi surfaces at the Fermi momentum $k_{\mathrm{F}}=n\pi
/2$ \cite{GGBOOK}, with filling factor $n=N/L$, where $N$ is the total
atomic number. At last, the chemical potential cuts the degenerate energy
bands, with the Fermi momenta $\pm k_{\mathrm{F}}$. As a result, the
momentum distribution function $n_{\uparrow }(k)$ has a plateau of $1$, with
sharp edges at $k=\pm k_{\mathrm{F}}$; see the black solid curve of Fig.~\ref%
{n_k}(a). In the large-$k$ limit, this plateau of $1$ disappears and two
plateaus of $0$ emerge, as expected. For the different chemical potentials,
the fillings are similar, and the momentum distributions are thus similar;
see the black solid curves of Figs.~\ref{n_k}(b) and~\ref{n_k}(c).

In the presence of SOC ($\lambda /t\neq 0$), the results are very
interesting. In this case, the energy bands governed by the Hamiltonian $%
H_{1}$ split into two nondegenerate subbands, whose minima and maxima are
given respectively by
\begin{equation}
k_{\mathrm{min}}=\pm 2\arctan (\frac{-t+\sqrt{t^{2}+\lambda ^{2}}}{\lambda }%
),  \label{kmin}
\end{equation}%
\begin{equation}
k_{\mathrm{max}}=\pm 2\arctan (\frac{t+\sqrt{t^{2}+\lambda ^{2}}}{\lambda }).
\label{kmax}
\end{equation}%
It is easy to find that $k_{\mathrm{min}}+k_{\mathrm{max}}=\pi $. For $%
\lambda =0$, $k_{\mathrm{min}}=0$, as expected. In the presence of SOC ($%
\lambda \neq 0$), there are generally four Fermi points, when the chemical
potential cuts two nondegenerate subbands. If the chemical potential is
identical to the critical chemical potential $\tilde{\mu}$ that just cuts
the contact point [see the blue curve in Fig.~\ref{energyband}(b)], three
Fermi points can emerge. More importantly, for the different chemical
potentials, the fillings are quite different; see Fig.~\ref{energyband}.
These different fillings affect dramatically on momentum distributions. As
examples, we plot, in Fig.~\ref{n_k}, the momentum distribution functions $%
n_{\uparrow }(k)$ for the different SOC strengths $\lambda /t$, when the
filling factors are chosen as (a) $n=0.3$, (b) $n=1$, and (c) $n=1.7$.

It can be seen from Fig.~\ref{n_k}(a) that for a smaller filling, any SOC
leads to a new momentum distribution, in which the corresponding function $%
n_{\uparrow }(k)$ has two plateaus of $1/2$, apart from the conventional
plateau of $1$. The physical explanation is given as follows. When the SOC
strength is not strong enough (see, for example, $\lambda /t=0.2$), $\mu >%
\tilde{\mu}$, and thus both the lower and upper subbands are partly
occupied; see Fig.~\ref{energyband}(c). The occupation in the upper subband
determines the plateau of $1$, while the occupation in the lower subband
governs two plateaus of $1/2$. With the increasing of the SOC strength $%
\lambda /t$, i.e., the chemical potential $\mu $ decreases \cite{ZQY11,GC12}%
, the occupation in the upper subband becomes less, and thus the wide of the
plateau of $1$ becomes narrower. In particular, when $\lambda /t=0.52$, the
plateau of $1$ disappears, since in this case $\mu =\tilde{\mu} $, and thus
no occupation in the upper subband can be found. If further increasing the
SOC strength $\lambda /t$, i.e., $\mu <\tilde{\mu}$, a plateau of $0$
emerges around $k=0$ (not in the large-$k$ limit). This is because there is
no occupation around $k=0$ when $\mu <\tilde{\mu}$; see Fig.~\ref{energyband}%
(a). For the half filling ($n=1$), we find that the momentum distribution
function $n_{\uparrow }(k)$ usually has two plateaus of $1$ and $1/2$ [see
Fig.~\ref{n_k}(b)], because $\mu >\tilde{\mu}$ and both the lower and upper
subbands are partly occupied. For a larger filling factor (see, for example,
$n=1.7$), the momentum distribution function $n_{\uparrow }(k)$ has two
plateaus of $1$ and $1/2$ when $\lambda /t=2.0$; see Fig.~\ref{n_k}(c). In
this case, $\mu \gg \tilde{\mu}$ and the lower subband is totally occupied
with a part occupation in the upper subband; see Fig.~\ref{energyband}(d).
This case cannot occur in the smaller filling factors. We call the
corresponding phase, with above unconventional momentum distributions, the
\textit{SOC-induced metallic phase}. Finally, in order to see clearly the
above evolution of momentum distribution with respect to the filling factor,
we plot the momentum distribution functions $n_{\uparrow }(k)$ for the
different filling factors $n$ in Fig.~\ref{n_k}(d).

It should be noticed that in the case with a Zeeman field, but without SOC,
the chemical potential cutting the Zeeman-split bands also results in more
than two Fermi points. But the momentum distribution function $n_{\uparrow
}(k)$ is quite different from that induced by SOC. In such case, the
momentum distribution function $n_{\uparrow }(k)$ only has the plateaus of $%
1 $. Moreover, for the different chemical potentials, the fillings are
similar, and thus the momentum distribution functions $n_{\uparrow }(k)$ are
also similar. However, in our considered case with SOC, but without the
Zeeman field, the momentum distribution function $n_{\uparrow }(k)$ has the
new plateaus of $1/2$, apart from $1$, and moreover, the filling-dependent
unconventional momentum distributions emerge.

\begin{figure}[t]
\centering\includegraphics[width = 3.5in]{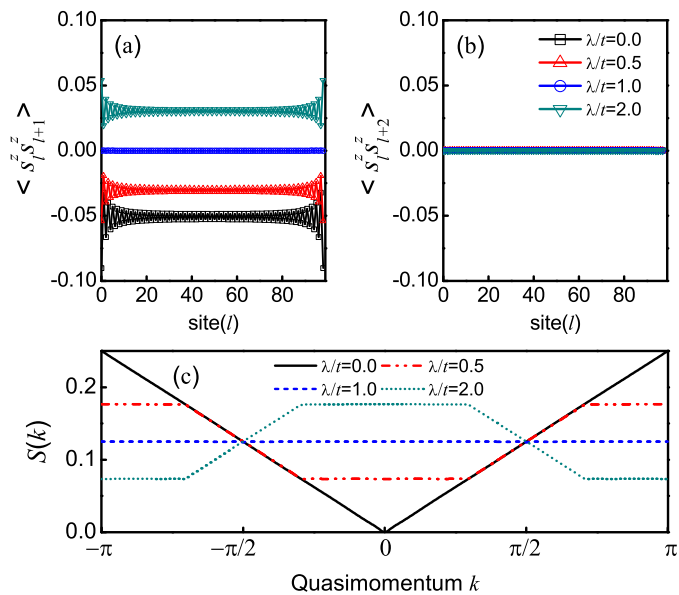}
\caption{The spin correlations between (a) the nearest-neighbor sites, i.e.,
$\left\langle s_{l}^{z}s_{l+1}^{z}\right\rangle $, and (b) the
next-nearest-neighbor sites, i.e., $\left\langle
s_{l}^{z}s_{l+2}^{z}\right\rangle $, for the different SOC strengths $%
\protect\lambda /t$. (c) The spin-structure factors $S(k)$ for the different
SOC strengths $\protect\lambda /t$. In all subfigures, the filling factor,
the on-site repulsive interaction strength, and the lattice length are given
by $n=1$, $U=0$, and $L=100$, respectively.}
\label{s_soc}
\end{figure}

\subsection{Spin-correlation function and spin-structure factor}

Since SOC can make Fermi atoms hop between the nearest-neighbor sites, with
spin flipping [see Fig.~\ref{setup}(a) and the Hamiltonian (\ref{HSOC})], it
has a competition with the conventional spin-independent hopping. This
competition has a strong effect on spin distributions of different sites. To
see this clearly, we consider the spin-correlation function in Eq.~(\ref{SCR}%
).

In Figs.~\ref{s_soc}(a) and~\ref{s_soc}(b), we plot the spin correlations
between the nearest-neighbor sites, i.e., $\left\langle
s_{l}^{z}s_{l+1}^{z}\right\rangle $, and the next-nearest-neighbor sites,
i.e., $\left\langle s_{l}^{z}s_{l+2}^{z}\right\rangle $, respectively. These
two subfigures show clearly that in the absence of SOC ($\lambda /t=0$), $%
\left\langle s_{l}^{z}s_{l+1}^{z}\right\rangle <0$ and $\left\langle
s_{l}^{z}s_{l+2}^{z}\right\rangle =0$, which means that this short-range
spin correlation is negative. In the presence of SOC ($\lambda /t\neq 0$),
the spin correlation is still short range. Interestingly, with the
increasing of the SOC strength $\lambda /t$, the spin correlation varies
from the negative to the positive at the critical point $\lambda _{c}/t=1$.
The physical reason will be illustrated in the following section.

In Fig.~\ref{s_soc}(c), we plot the spin-structure factors $S(k)$ for the
different SOC strengths $\lambda /t$ at the half filling ($n=1$). In the
absence of SOC ($\lambda /t=0$), the spin-structure factor $S(k)$ evolves as
a straight line from $0$ to $0.25$, and has a cusp at $k=\pm \pi $ \cite%
{MO90}. For a finite SOC strength $\lambda /t=0.5$, the spin-structure
factor is $S(0)>0$ and $0<S(\pi )<0.25$. When $\lambda /t=1$, the
spin-structure factor $S(k)$ becomes a constant $0.125$, since in this case
no short-range spin correlation can be found. When $\lambda /t=2$, the
spin-structure factor $S(k)$ is similar to the case of $\lambda /t=0.5$.

\begin{figure}[t]
\centering\includegraphics[width = 6.0in]{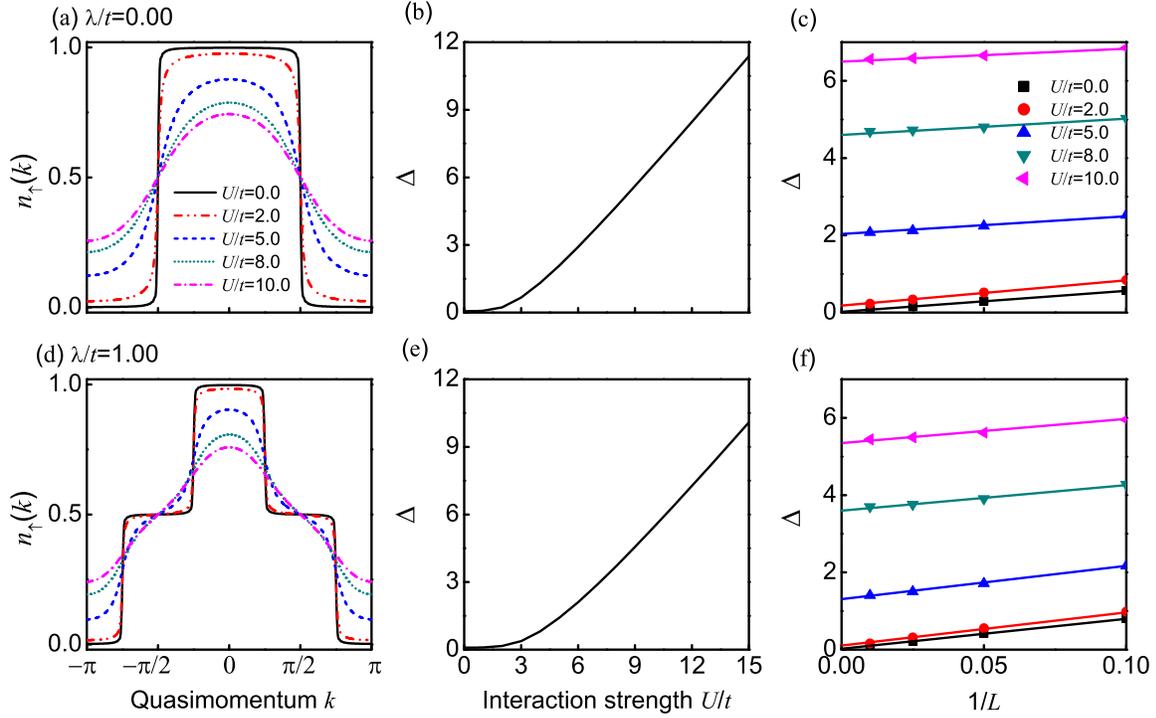}
\caption{The momentum distribution functions $n_{\uparrow }(k)$ for the
different on-site repulsive interaction strengths $U/t$ (Left panel), the
gap $\Delta $ as a function of $U/t$ (Center panel), and the gaps $\Delta $
as functions of $1/L$\ for the different $U/t$ (Right panel) at the half
fillings ($n=1$). In all subfigures, the SOC strength is chosen as (a-c) $%
\protect\lambda /t=0$ and (d-f) $\protect\lambda /t=1.0$. In (a), (b), (d),
and (e), the lattice length is given by $L=100$. The scaling behaviors in
(c) and (f) show that in the thermodynamical limit ($L\rightarrow \infty $),
$\Delta \equiv 0$ for $U/t=0$. }
\label{u_n_k}
\end{figure}

\section{With on-site repulsive interaction}

We now consider the case with the on-site repulsive interaction ($U/t>0$).
In the absence of SOC ($\lambda /t=0$), it has been demonstrated exactly
that for the 1D homogeneous Fermi-Hubbard model at the half filling ($n=1$),
the metallic phase occurs at $U=0$. For $U>0$, the system is always located
at the antiferromagnetic Mott insulator \cite{EHL68}, in which the spin
orientations of the nearest-neighbor sites are antiparallel \cite{MO90}. It
means that a second-order quantum phase transition between the metallic
phase and the antiferromagnetic Mott insulator emerges at $U=0$. In the
following, we will show that SOC can drive the system from the
antiferromagnetic to ferromagnetic Mott insulators with spin rotating (i.e.,
the spin orientations of the nearest-neighbor sites are not parallel or
antiparallel), and predict a second-order quantum phase transition between
the spin-rotating ferromagnetic Mott insulator and the SOC-induced metallic
phase at the half filling ($n=1$).

\begin{figure}[t]
\centering\includegraphics[width = 6.0in]{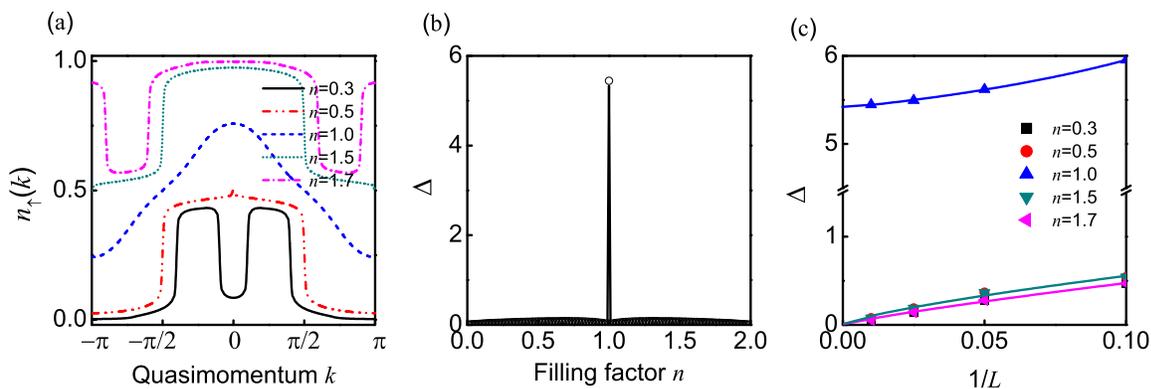}
\caption{(a) The momentum distribution functions $n_{\uparrow }(k)$ for the
different filling factors $n$. (b) The gap $\Delta $ as a function of the
filling factor $n$. (c) The gap $\Delta $ as functions of $1/L$ for the
different filling factors $n$. In all subfigures, the SOC strength and the
on-site repulsive interaction strength are given by $\protect\lambda /t=1.0$
and $U/t=10$, respectively. In both (a) and (b), the lattice length is given
by $L=100$. The scaling behavior in (c) shows that in the thermodynamical
limit ($L\rightarrow \infty $), $\Delta \equiv 0$, except for the half
filling ($n=1$).}
\label{n_k_n}
\end{figure}

\subsection{Momentum distribution function and gap}

Figures~\ref{u_n_k}(a) and \ref{u_n_k}(d) show the momentum distribution
functions $n_{\uparrow }(k)$ for the different SOC strengths $\lambda /t$ at
the half filling ($n=1$). In the absence of SOC ($\lambda /t=0$) and the
on-site repulsive interaction strength ($U/t=0$), the momentum distribution
function $n_{\uparrow }(k)$ has two sharp edges at $k=\pm k_{\mathrm{F}}$;
see the black solid curve in Fig.~\ref{u_n_k}(a). When increasing the
on-site repulsive interaction strength $U/t$, the momentum distribution
functions $n_{\uparrow }(k)$ become smoother and all sharp edges disappear.
In order to see the relevant physics more clearly, we introduce the gap \cite%
{EHL68}
\begin{equation}
\Delta =\mu ^{+}-\mu ^{-},  \label{gap}
\end{equation}%
where $\mu ^{+}=E_{g}(N+1)-E_{g}(N)$ and $\mu ^{-}=E_{g}(N)-E_{g}(N-1)$,
with the ground-state energy $E_{g}$. This gap reflects the difference
between the energy required to add ($\mu ^{+}$) and remove ($\mu ^{-}$) a
Fermi atom from the ground state. In the thermodynamic limit ($L\rightarrow
\infty $), it has been demonstrated rigorously that $\Delta \equiv 0$ for $%
U/t=0$ and $\Delta \neq 0$ for $U/t>0$. Moreover, $\Delta \equiv 0$
corresponds to the metallic phase and $\Delta \neq 0$ corresponds to the
Mott insulator \cite{EHL68}. In our numerical results, due to finite-size
effects, the gap is not absolute zero when $U/t=0$; see Fig.~\ref{u_n_k}(b).
However, this result can be extrapolated to the thermodynamic limit by a
finite-size-scaling analysis \cite{SD07,KB07}. As shown in Fig.~\ref{u_n_k}%
(c), we find $\Delta \equiv 0$ for $U/t=0$ and $\Delta \neq 0$ for $U/t>0$
in the thermodynamical limit. Therefore, here we call the phase, in which $%
n_{\uparrow }(k)$ has sharp edges at $k=\pm k_{\mathrm{F}}$ and $\Delta
\equiv 0$ (thermodynamical limit), the metallic phase, whereas the phase, in
which $n_{\uparrow }(k)$ become smoother, all sharp edges disappear, and $%
\Delta \neq 0$ (thermodynamical limit), is referred as the Mott insulator
\cite{EHL68}. Since in this Mott insulator the spin orientations of the
nearest-neighbor sites are antiparallel, the phase is finally called the
antiferromagnetic Mott insulator (see also the following discussions).

In the presence of SOC (see, for example, $\lambda /t=1$), the momentum
distribution function $n_{\uparrow }(k)$ is unconventional and $\Delta
\equiv 0$ (thermodynamical limit), when $U/t=0$; see the black solid curve
in Figs.~\ref{u_n_k}(d)-\ref{u_n_k}(f). It implies that the system is
located at the SOC-induced metallic phase. When increasing the on-site
repulsive interaction strength $U/t$, these momentum distribution functions $%
n_{\uparrow }(k)$ also become smoother, all sharp edges also disappear, and $%
\Delta \neq 0$ (thermodynamical limit), i.e., the system enters into the
Mott insulator. However, as will be shown in the next subsection, these Mott
insulators, without SOC or with SOC, are quite different. Without SOC, the
Mott insulator is antiferromagnetic, whereas it becomes spin-rotating
antiferromagnetic if $0<\lambda/t<1$, and spin-rotating ferromagnetic if $%
\lambda/t>1$.

\begin{figure}[t]
\centering\includegraphics[width = 3.5in]{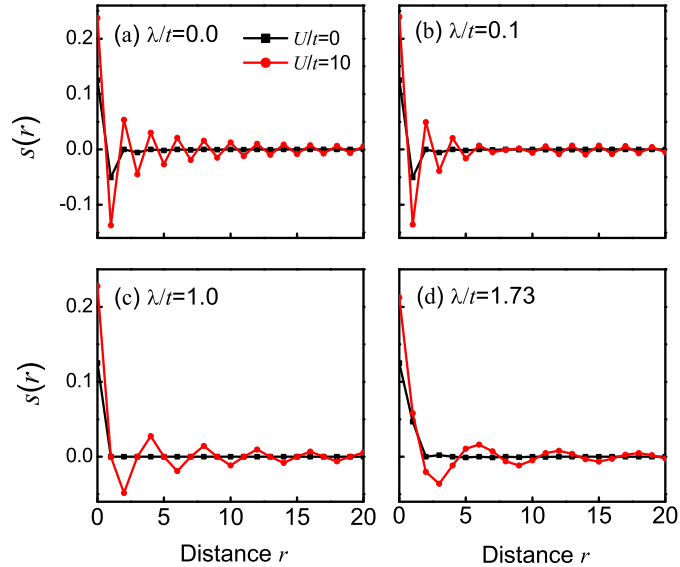}
\caption{The spin-correlation functions $s(r)$ for the different SOC
strengths (a) $\protect\lambda /t=0$, (b) $\protect\lambda /t=0.1$, (c) $%
\protect\lambda /t=1$, and (d) $\protect\lambda /t=1.73$ at the half filling
($n=1$). In all subfigures, the lattice length is given by $L=100$.}
\label{u_s_r}
\end{figure}

In Figs.~\ref{n_k_n}(a)-\ref{n_k_n}(c), we plot the momentum distribution
functions $n_{\uparrow }(k)$ and the gap $\Delta $ for the different filling
factors $n$, when $\lambda /t=1$ and $U/t=10.0$. If $n\neq 1$, all sharp
edges (Fermi points) in the momentum distribution functions $n_{\uparrow
}(k) $ still exist and the zero gap, $\Delta \equiv 0$ (thermodynamical
limit), remains. These indicate that the system is always located at the
SOC-induced metallic phase for any on-site repulsive interaction strength $%
U/t$. However, at the half filling ($n=1$), the momentum distribution
function $n_{\uparrow }(k)$ becomes smoother and $\Delta $ $\neq 0 $
(thermodynamical limit) for any on-site repulsive interaction strength $U/t$%
. These mean that these Mott insulators only occur at the half filling ($n=1$%
), which is similar to the result without SOC \cite{EHL68}.

\begin{figure}[t]
\centering\includegraphics[width = 4.0in]{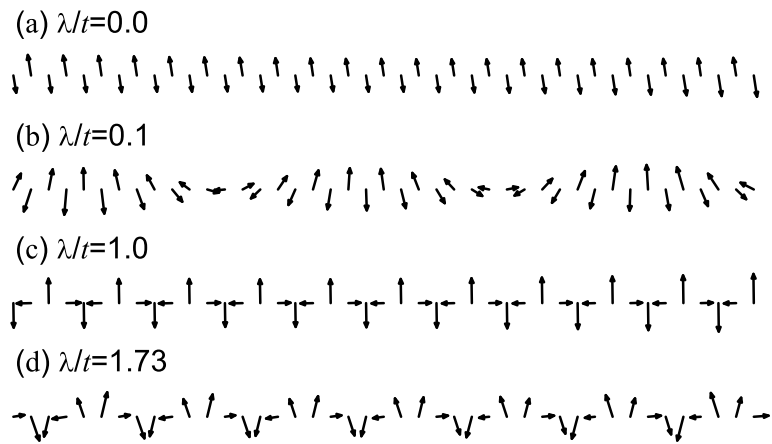}
\caption{Vector plots of DMRG results $s_{l}=(s_{l}^{x},s_{l}^{z})$ for the
different SOC strengths $\protect\lambda /t$ at the half filling ($n=1$). In
all subfigures, the on-site repulsive interaction strength and the lattice
length are given respectively as $U/t=10$ and $L=100$, which are the same
values as the red curve of Fig.~\protect\ref{u_s_r}.}
\label{lattice}
\end{figure}

\subsection{Spin-correlation function and spin-structure factor}

In Figs.~\ref{u_s_r}(a)-\ref{u_s_r}(d), we plot the spin-correlation
functions $s(r)$ for the different SOC strengths $\lambda /t$. These figures
show clearly that in the presence of the on-site repulsive interaction ($%
U/t\neq 0$), quasi-long-range spin correlation appears, i.e., $s(r>1)\neq 0$%
. This is contrast to the result without the on-site repulsive interaction,
in which only the short-range spin correlation emerges; see Figs.~\ref{s_soc}%
(a) and~\ref{s_soc}(b). When increasing the SOC strength $\lambda /t$, the
spin-correlation functions between the nearest-neighbor sites vary from the
negative to the positive. In order to see this physics more clearly, we
plot, in Figs.~\ref{lattice}(a)-\ref{lattice}(d), spin distributions of each
sites, i.e., $s_{l}=(s_{l}^{x},s_{l}^{z})$, where $s_{l}^{x}=\left\langle
G\right\vert c_{l\uparrow }^{\dagger }c_{l\downarrow }+c_{l\downarrow
}^{\dagger }c_{l\uparrow }\left\vert G\right\rangle $ and $%
s_{l}^{z}=\left\langle G\right\vert c_{l\uparrow }^{\dagger }c_{l\uparrow
}-c_{l\downarrow }^{\dagger }c_{l\downarrow }\left\vert G\right\rangle $
with the ground-state wavefunction $\left\vert G\right\rangle $, for the
different SOC strengths $\lambda /t$. We define an intersection angle $%
\theta $ between the different spin orientations of the nearest-neighbor
sites.

In the absence of SOC ($\lambda /t=0$), the spin orientations of the
nearest-neighbor sites are antiparallel and $\theta =\pi $; see Figs.~\ref%
{lattice}(a). Moreover, the spin-correlation function $s(r)>0$ if $r$ is
odd, while $s(r)<0$ if $r$ is even; see Fig.~\ref{u_s_r}(a). These mean that
the corresponding spin-spin interactions of the nearest-neighbor sites are
antiferromagnetic. In addition, the spin-correlation function $s(r)$ decays
as a power law and changes the sign with a period $T=2$ \cite{MR04}; see
also Fig.~\ref{u_s_r}(a). This phase is usually called the antiferromagnetic
Mott insulator. In the presence of SOC (see, for example, $\lambda /t=0.1$),
the spin orientations of the nearest-neighbor sites are not antiparallel
(i.e., the spins are rotating) and $\pi /2<\theta <\pi $; see Figs.~\ref%
{lattice}(b). Since in this case the traditional spin-independent hopping
still plays a dominate role, the quasi-long-range antiferromagnetic spin
correlation remains, but with a period $2<T<4$; see Fig.~\ref{u_s_r}(b).
Thus, we call the corresponding phase the spin-rotating antiferromagnetic
Mott insulator. When $\lambda /t=1$, the spin orientations of the
nearest-neighbor sites are vertical and $\theta =\pi /2$; see Figs.~\ref%
{lattice}(c). Moreover, $s(1)=s(3)=\cdots =0$ with a period $T=4$, which
means that no spin correlation between the nearest-neighbor sites can be
found; see Fig.~\ref{u_s_r}(c). At the strong SOC ($\lambda /t>1$), the
SOC-induced hopping plays a dominate role. In this case, the spin
orientations of the nearest-neighbor sites tend to parallel and $\theta <\pi
/2$; see Figs.~\ref{lattice}(d). Moreover, $s(1)>0$, which indicates that
the spin-spin interactions of the nearest-neighbors sites become
ferromagnetic. In addition, the quasi-long-range spin-correlation function $%
s(r)$ also decays as a power law, but changes the sign with a period $T>4$;
see Fig.~\ref{u_s_r}(d). We call the corresponding phase the spin-rotating
ferromagnetic Mott insulator. From above discussions, we argue that SOC can
drive the system from the spin-rotating antiferromagnetic Mott insulator to
the spin-rotating ferromagnetic Mott insulator.

\begin{figure}[t]
\centering\includegraphics[width = 4.0in]{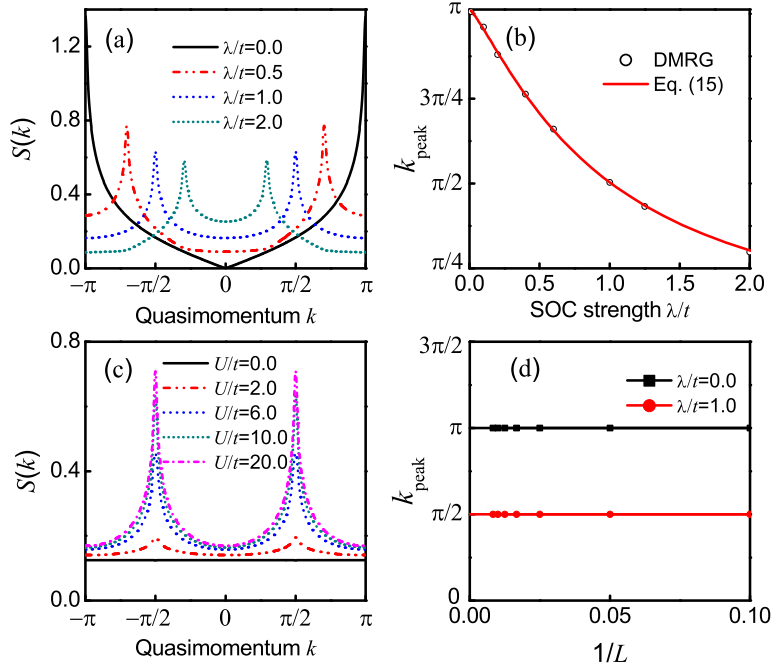}
\caption{(a) The spin-structure factors $S(k)$ for the different SOC
strengths $\protect\lambda /t$ as functions of the quasimomentum $k$. (b)
The momentum $k_{\mathrm{peak}}$ as a function of the SOC strength $\protect%
\lambda /t$. (c) The spin-structure factors $S(k)$ for the different on-site
repulsive interaction strengths $U/t$ as functions of the quasimomentum $k$,
when the SOC strength is given by $\protect\lambda /t=1$. In (a)-(c), the
lattice length is given by $L=100$. (d) The momentum $k_{\mathrm{peak}}$ as
a function of $1/L$. In (a), (b), and (d), the on-site repulsive interaction
strength is given by $U/t=10$. In all subfigures, the filling factor is
given by $n=1$.}
\label{u_s_k_u=10_n=1_soc}
\end{figure}

Figure~\ref{u_s_k_u=10_n=1_soc}(a) shows the experimentally-measurable
spin-structure factors $S(k)$ for the different SOC strengths $\lambda /t$
at the half filling ($n=1$). In the absence of SOC ($\lambda /t=0$), the
system has the antiferromagnetic order \cite{MO90,JEH83}, and the
spin-structure factor $S(k)$ has a peak at the momentum $k_{\mathrm{peak}%
}=\pi $, as expected. When increasing the SOC strength $\lambda /t$, the
peak still exists and varies as%
\begin{equation}
k_{\mathrm{peak}}=2k_{\max }-\pi =4\arctan (\frac{t+\sqrt{t^{2}+\lambda ^{2}}%
}{\lambda })-\pi .  \label{kpeak}
\end{equation}%
It is easy to find from Eq.~(\ref{kpeak}) that when $\lambda /t=1$, $k_{%
\mathrm{peak}}=\pi /2$; see Fig.~\ref{u_s_k_u=10_n=1_soc}(b). When $%
0<\lambda /t<1$, the system has the spin-rotating antiferromagnetic Mott
insulator with $\pi /2<k_{\mathrm{peak}}<\pi $. When $1<\lambda /t<2$, the
system has the spin-rotating ferromagnetic Mott insulator with $k_{\mathrm{%
peak}}<\pi /2$. Since the antiferromagnetic order has been detected
experimentally by measuring $k_{\mathrm{peak}}$ via the time-of-flight
imaging \cite{RAH15}, these spin-rotating ferromagnetic and
antiferromagnetic orders can also be detected by the same method. For a
fixed SOC strength $\lambda /t=1$, when increasing the on-site repulsive
interaction strength $U/t$, the momentum $k_{\mathrm{peak}}$ remains
unchanged, while the magnitudes of peaks increase; see Fig.~\ref%
{u_s_k_u=10_n=1_soc}(c). In Fig.~\ref{u_s_k_u=10_n=1_soc}(d), we present a
finite-size-scaling analysis of the momentum $k_{\mathrm{peak}}$. This
figure shows that the momentum $k_{\mathrm{peak}}$ remains unchanged when
increasing the lattice length $L$.

\subsection{Phase diagram}

\begin{figure}[t]
\centering\includegraphics[width = 3.0in]{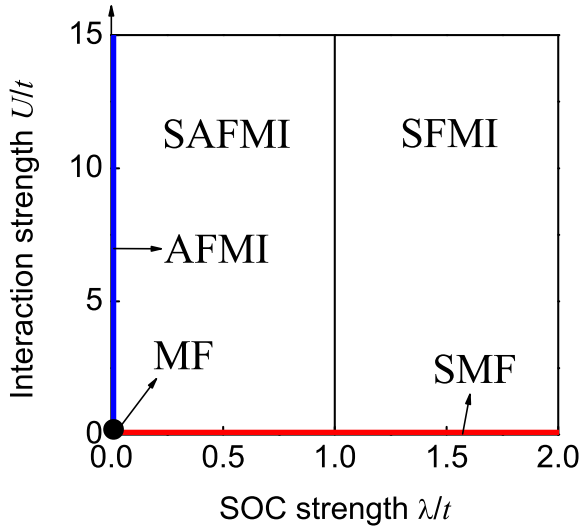}
\caption{Schematic phase diagram as a function of the SOC strength $\protect%
\lambda /t$ and the on-site repulsive interaction strength $U/t$ at the half
filling ($n=1$). The black dot denotes the metallic phase (MF), the blue
line denotes the antiferromagnetic Mott insulator (AFMI), and the red line
denotes the SOC-induced metallic phase (SMF). In the other abbreviations,
SAFMI and SFMI denote the spin-rotating antiferromagnetic Mott insulator and
the spin-rotating ferromagnetic Mott insulator, respectively.}
\label{phase}
\end{figure}

Based on the predicted properties of the momentum distribution, the gap, and
the spin correlation, we find that the homogeneous Hamiltonian (\ref{TH})
has five phases, including the metallic phase, the SOC-induced metallic
phase, the antiferromagnetic Mott insulator, the spin-rotating
antiferromagnetic Mott insulator, and the spin-rotating ferromagnetic Mott
insulator, at the half filling ($n=1$). In the metallic phase, the system
has conventional momentum distribution with the gap $\Delta \equiv 0$
(thermodynamical limit), while it becomes filling-dependent unconventional
momentum distribution but with the same zero gap (thermodynamical limit) in
the SOC-induced metallic phase. In the antiferromagnetic, spin-rotating
antiferromagnetic, and spin-rotating ferromagnetic Mott insulators, the
system has a nonzero gap (thermodynamical limit) and the quasi-long-range
spin correlation decaying as a power law. However, in the antiferromagnetic
Mott insulator, $\theta =\pi $, $T=2$, and $k_{\mathrm{peak}}=\pi $. In the
spin-rotating antiferromagnetic Mott insulator, $\pi /2<\theta <\pi $, $%
2<T<4 $, and $\pi /2<k_{\mathrm{peak}}<\pi $. Whereas in the spin-rotating
ferromagnetic Mott insulator, $\theta <\pi /2$, $T>4$, and $k_{\mathrm{peak}%
}<\pi /2$.

In Fig.~\ref{phase}, we give a schematic phase diagram as a function of the
SOC strength $\lambda /t$ and the on-site repulsive interaction strength $%
U/t $. In this figure, the Mott insulator and the metallic phase are
separated by the gap $\Delta $ (thermodynamical limit) \cite{EHL68}. The
metallic phase and the SOC-induced metallic phase are separated by the
momentum distribution function $n_{\uparrow }(k)$. The antiferromagnetic
Mott insulator and the spin-rotating antiferromagnetic Mott insulator are
separated by $\theta =\pi $, $T=2$, and $k_{\mathrm{peak}}=\pi $. The
spin-rotating antiferromagnetic Mott insulator and the spin-rotating
ferromagnetic Mott insulator are separated by $\theta =\pi /2$, $T=4$, and $%
k_{\mathrm{peak}}=\pi /2$.

We also find that the transitions between the metallic phase and the
antiferromagnetic Mott insulator, between the SOC-induced metallic phase and
the spin-rotating antiferromagnetic Mott insulator, and between the
SOC-induced metallic phase and the spin-rotating ferromagnetic Mott
insulator are of second order, because the second-order derivative of the
ground-state energy is discontinuous at the critical points. For the other
transitions, the ground-state energy and its derivative vary smoothly. It
means that no phase transition can be found, although for the different
phases, the momentum distribution and the spin correlation are different.

\section{Discussions and conclusions}

Before ending up this paper, we make one remark. In the real experiment, the
Zeeman field usually exists. When the Zeeman field is considered, an extra
Hamiltonian
\begin{equation}
H_{_{\mathrm{Zeeman}}}=h\sum_{l}(n_{l\uparrow }-n_{l\downarrow })
\label{Zem}
\end{equation}%
should be added in the Hamiltonian (\ref{TH}). This Zeeman field can lead to
spin flipping in the same site. Moreover, it has a strong competition with
SOC, which makes Fermi atoms hop between the nearest-neighbor sites with
spin flipping, and the conventional spin-independent hopping between the
nearest-neighbor sites. As a consequence, rich magnetic properties can
emerge. The deep understanding of relevant behavior is very complicated (we
need introduce more physical quantities), but interesting. We leave this
important problem for further investigation.

In summary, we have investigated spin-orbit coupled repulsive Fermi atoms in
a 1D optical lattice by the DMRG method. We have found that SOC can generate
the filling-dependent unconventional momentum distribution, whose
corresponding phase is called the SOC-induced metallic phase, and drive the
system from the spin-rotating antiferromagnetic Mott insulator to the
spin-rotating ferromagnetic Mott insulator. We have predicted a second-order
quantum phase transition between the spin-rotating ferromagnetic Mott
insulator and the SOC-induced metallic phase at the strong SOC. Finally, we
have also shown that the momentum, at which peak of the spin-structure
factor appears, can be affected dramatically by SOC. The analytical
expression of this momentum with respect to the SOC strength has also been
derived. Attributed to the recent experimental realization of the spin-orbit
coupled Bose-Einstein condensates in the 1D optical lattice \cite{CH14}, we
expect that our predictions could be observed in the near future. In
particular, the spin-rotating ferromagnetic and antiferromagnetic
correlations can be detected by measuring the SOC-dependent spin-structure
factor via the time-of-flight imaging \cite{RAH15}.

\ack

We acknowledge Professors Ming Gong and Xiaoling Cui for their valuable
discussions. This work is supported in part by the 973 program under Grant
No.~2012CB921603; the NNSFC under Grant No.~11422433, No.~11434007, and
No.~61275211; the PCSIRT under Grant No.~IRT13076; the NCET under Grant
No.~13-0882; the FANEDD under Grant No.~201316; OYTPSP; and SSCC.


\section*{References}

\begin{thebibliography}{99}
\bibitem{MK05} K\"{o}hl M, Moritz H, S\"{o}tferle T, G\"{u}nter K and
Esslinger T 2005 \textit{Phys. Rev. Lett.} \textbf{94} 080403

\bibitem{HM05} Moritz H, S\"{o}tferle T, G\"{u}nter K, K\"{o}hl M and
Esslinger T 2005 \textit{Phys. Rev. Lett.} \textbf{94} 210401

\bibitem{JKC06} Chin J K, Miller D E, Liu Y, Stan C, Setiawan W, Sanner C,
Xu K and Ketterle W 2006 \textit{Nature} \textbf{443} 961

\bibitem{TR06} Rom T, Best Th, Oosten D van, Schneider U, F\"{o}lling S,
Paredes B and Bloch I 2006 \textit{Nature} \textbf{444} 733

\bibitem{TS06} St\"{o}ferle T, Moritz H, G\"{u}nter K, K\"{o}hl M and
Esslinger T 2006 \textit{Phys. Rev. Lett.} \textbf{96} 030401

\bibitem{NS07} Strohmaier N, Takasu Y, G\"{u}nter K, J\"{o}rdens R, K\"{o}hl
M, Moritz H and Esslinger T 2007 \textit{Phys. Rev. Lett.} \textbf{99} 220601

\bibitem{RJ08} J\"{o}rdens R, Strohmaier N, G\"{u}nter K, Moritz H and
Esslinger T 2008 \textit{Nature} \textbf{455} 204

\bibitem{US08} Schneider U, Hackerm\"{u}ller L, Will S, Best T, Bloch I,
Costi T A, Helmes R W, Rasch D and Rosch A 2008 \textit{Science} \textbf{322}
1520

\bibitem{YAL10} Liao Y-a, Rittner A S C, Paprotta T, Li W, Partridge G B,
Hulet R G, Baur S K and Mueller E J 2010 \textit{Nature} \textbf{467} 567

\bibitem{NS10} Strohmaier N, Greif D, J\"{o}rdens R, Tarruell L, Moritz H,
Esslinger T, Sensarma R, Pekker D, Altman E and Demler E 2010 \textit{Phys.
Rev. Lett.} \textbf{104} 080401

\bibitem{LT12} Tarruell L, Greif D, Uehlinger T, Jotzu G and Esslinger T
2012 \textit{Nature} \textbf{483} 302

\bibitem{JSK12} Krauser J S, Heinze J, Fl\"{a}schner N, G\"{o}tze S, J\"{u}%
rgensen O, L\"{u}hmann D, Becker C and Sengstock K 2012 \textit{Nat. Phys.}
\textbf{8} 813

\bibitem{DG13} Greif D, Uehlinger T, Jotzu G, Tarruell L and Esslinger T
2013 \textit{Science} \textbf{340} 1307

\bibitem{JH13} Heinze J, Krauser J S, Fl\"{a}schner N, Hundt B, G\"{o}tze S,
Itin A, Mathey L, Sengstock K and Becker C 2013 \textit{Phys. Rev. Lett.}
\textbf{110} 085302

\bibitem{GP14} Pagano G, Mancini M, Cappellini G, Lombardi P, Sch\"{a}fer F,
Hu H, Liu X-J, Catani J, Sias C, Inguscio M and Fallani L 2014 \textit{Nat.
Phys.} \textbf{10} 198

\bibitem{PMD15} Duarte P M, Hart R A, Yang T-L, Liu X-X, Paiva T, Khatami E,
Scalettar R T, Trivedi N and Hulet R G 2015 \textit{Phys. Rev. Lett.}
\textbf{114} 070403

\bibitem{MR03} Rigol M, Muramatsu A, Batrouni G G and Scalettar R T 2003
\textit{Phys. Rev. Lett.} \textbf{91} 130403

\bibitem{MR04} Rigol M and Muramatsu A 2004 \textit{Phys. Rev A.} \textbf{69}
053612

\bibitem{XJ05} Liu X-J, Drummond P D and Hu H 2005 \textit{Phys. Rev. Lett.}
\textbf{94} 136406

\bibitem{MI05} Iskin M and S\'{a} de Melo C A R 2005 \textit{Phys. Rev. B}
\textbf{72} 224513

\bibitem{MI130} Iskin M 2013 \textit{Phys. Rev. A} \textbf{88} 053606

\bibitem{EZ06} Zhao E and Paramekanti A 2006 \textit{Phys. Rev. Lett.}
\textbf{97} 230404

\bibitem{XG07} Gao X, Rizzi M, Polini M, Fazio R, Tosi M, Campo V and
Capelle K 2007 \textit{Phys. Rev. Lett.} \textbf{98} 030404

\bibitem{LM07} Mathey L, Tsai S-W and Castro Neto A H 2007 \textit{Phys.
Rev. B} \textbf{75} 174516

\bibitem{AM07} Moreo A and Scalapino D 2007 \textit{Phys. Rev. Lett.}
\textbf{98} 216402

\bibitem{SJG07} Gu S-J, Fan R and Lin H-Q 2007 \textit{Phys. Rev. B} \textbf{%
76} 125107

\bibitem{FK07} Karim Pour F, Rigol M, Wessel S and Muramatsu A 2007 \textit{%
Phys. Rev. B} \textbf{75} 161104

\bibitem{BA07} Andersen B and Bruun G 2007 \textit{Phys. Rev. A} \textbf{76}
041602

\bibitem{MRB08} Bakhtiari M R, Leskinen M J and T\"{o}rm\"{a} P 2008 \textit{%
Phys. Rev. Lett.} \textbf{101} 120404

\bibitem{MT08} Tezuka M and Ueda M 2008 \textit{Phys. Rev. Lett.} \textbf{100%
} 110403

\bibitem{MT10} Tezuka M and Ueda M 2010 \textit{New J. Phys.} \textbf{12}
055029

\bibitem{MM08} Machida M, Okumura M, Yamada S, Deguchi T, Ohashi Y and
Matsumoto H 2008 \textit{Phys. Rev. B} \textbf{78} 235117

\bibitem{KW08} Wu K and Zhai H 2008 \textit{Phys. Rev. B} \textbf{77} 174431

\bibitem{YC09} Chen Y, Wang Z, Zhang F and Ting C 2009 \textit{Phys. Rev. B}
\textbf{79} 054512

\bibitem{AH10} Chen A-H and Gao X 2010 \textit{Phys. Rev. A} \textbf{81}
013628

\bibitem{KS11} Sun K, Vincent Liu W, Hemmerich A and Das Sarma S 2011
\textit{Nat. Phys.} \textbf{8} 67

\bibitem{AY11} Yamamoto A, Yamada S, Okumura M and Machida M 2011 \textit{%
Phys. Rev. A} \textbf{84} 043642

\bibitem{KK12} Kobayashi K, Okumura M, Ota Y, Yamada S and Machida M 2012
\textit{Phys. Rev. Lett.} \textbf{109} 235302

\bibitem{DV12} Vol\v{c}ko D and Quader K F 2012 \textit{Phys. Rev. Lett.}
\textbf{109} 235303

\bibitem{ZS12} Shen Z, Radzihovsky L and Gurarie V 2012 \textit{Phys. Rev.
Lett.} \textbf{109} 245302

\bibitem{ST14} Wang S-T, Deng D-L and Duan L-M 2014 \textit{Phys. Rev. Lett.}
\textbf{113} 033002

\bibitem{ML07} Lewenstein M, Sanpera A, Ahufinger V, Damski B, Sen(De) A and
Sen U 2007 \textit{Adv. Phys.} \textbf{56} 243

\bibitem{TS10} Esslinger T 2010 \textit{Annu. Rev. Cond. Mat. Phys.} \textbf{%
1} 129

\bibitem{CC10} Chin C, Grimm R, Julienne P and Tiesinga E 2010 \textit{Rev.
Mod. Phys.} \textbf{82} 1225

\bibitem{EHL68} Lieb E H and Wu F Y 1968 \textit{Phys. Rev. Lett.} \textbf{20%
} 1445

\bibitem{PW12} Wang P, Yu Z-Q, Fu Z, Miao J, Huang L, Chai S, Zhai H and
Zhang J 2012 \textit{Phys. Rev. Lett.} \textbf{109} 095301

\bibitem{RAW13} Williams R A, Beeler M C, LeBlanc L J, Jim\'{e}nez-Garc\'{\i}%
a K and Spielman I B 2013 \textit{Phys. Rev. Lett.} \textbf{111} 095301

\bibitem{ZF14} Fu Z, Huang L, Meng Z, Wang P, Zhang L, Zhang S, Zhai H,
Zhang P and Zhang J 2014 \textit{Nat. Phys.} \textbf{10} 110

\bibitem{LWC12} Cheuk L W, Sommer A T, Hadzibabic Z, Yefsah T, Bakr W S and
Zwierlein M W 2012 \textit{Phys. Rev. Lett.} \textbf{109} 095302

\bibitem{ZY03} Zhang Y, Chen G and Zhang C 2013 \textit{Sci. Rep.} \textbf{3}
1937

\bibitem{KJ14} Jim\'{e}nez-Garci\'{a} K, LeBlanc L J, Williams R A, Beeler M
C, Qu C, Gong M, Zhang C and Spielman I B 2015 \textit{Phys. Rev. Lett.}
\textbf{114} 125301

\bibitem{MG11} Gong M, Tewari S and Zhang C 2011 \textit{Phys. Rev. Lett.}
\textbf{107} 195303

\bibitem{JZ11} Zhou J, Zhang W and Yi W 2011 \textit{Phys. Rev. A} \textbf{84%
} 063603

\bibitem{MG12} Gong M, Chen G, Jia S and Zhang C 2012 \textit{Phys. Rev.
Lett.} \textbf{109} 105302

\bibitem{KS12} Seo K, Han L and S\'{a} de Melo C A R 2012 \textit{Phys. Rev.
Lett.} \textbf{109} 105303

\bibitem{RW12} Wei R and Mueller E J 2012 \textit{Phys. Rev. A} \textbf{86}
063604

\bibitem{XIJL12} Liu X-J and Hu H 2012 \textit{Phys. Rev. A} \textbf{85}
033622

\bibitem{MI13} Iskin M and Suba\c{s}i A L 2013 \textit{Phys. Rev. A} \textbf{%
87} 063627

\bibitem{HH13} Hu H, Jiang L, Pu H, Chen Y and Liu X-J 2013 \textit{Phys.
Rev. Lett.} \textbf{110} 020401

\bibitem{CQU13} Qu C, Zheng Z, Gong M, Xu Y, Mao L, Zou X, Guo G and Zhang C
2013 \textit{Nat. Commun.} \textbf{4} 2710

\bibitem{WZ13} Zhang W and Yi W 2013 \textit{Nat. Commun.} \textbf{4} 3710

\bibitem{XJL13} Liu X-J and Hu H 2013 \textit{Phys. Rev. A} \textbf{88}
023622

\bibitem{LPG01} Gor'kov L P and Rashbar E I 2001 \textit{Phys. Rev. Lett.}
\textbf{87} 037004

\bibitem{CWZ08} Zhang C, Tewari S, Lutchyn R M and Sarma S D 2008 \textit{%
Phys. Rev. Lett.} \textbf{101} 160401

\bibitem{Sato09} Sato M, Takahashi Y and Fujimoto S 2009 \textit{Phys. Rev.
Lett.} \textbf{103} 020401

\bibitem{SSZ13} Zhang S-S, Yu X-L, Ye J and Liu W-M 2013 \textit{Phys. Rev. A%
} \textbf{87} 063623

\bibitem{SSZ14} Zhang S-S, Ye J and Liu W-M 2014 Itinerant ferromagnetism in
repulsively interacting spin-orbit coupled Fermi gas arXiv:1403.7031

\bibitem{XC14} Cui X and Ho T-L 2014 \textit{Phys. Rev. A} \textbf{89} 013629

\bibitem{CH14} Hamner C, Zhang Y, Khamehchi M A, Davis M J and Engels P 2015
\textit{Phys. Rev. Lett.} \textbf{114} 070401

\bibitem{XWG13} Guan X-W, Batchelor M T and Lee C 2013 \textit{Rev. Mod.
Phys.} \textbf{85} 1633

\bibitem{LJJ14} Liang J-J, Zhou X-F, Chui P-H, Zhang K, Gu S-J, Gong M, Chen
G and Jia S-T 2014 Spin-orbit coupling induced unconventional pairings in a
one-dimensional lattice arXiv:1404.3009

\bibitem{YHC14} Chan Y H 2015 \textit{Phys. Rev. B} \textbf{91} 235136

\bibitem{US05} Schollw\"{o}ck U 2005 \textit{Rev. Mod. Phys.} \textbf{77} 259

\bibitem{RAH15} Hart R A, Duarte P M, Yang T-L, Liu X, Paiva T, Khatami E,
Scalettar R T, Trivedi N, Huse D A and Hulet R G 2015 \textit{Nature}
\textbf{519} 211

\bibitem{MO98} Olshanii M 1998 \textit{Phys. Rev. Lett.} \textbf{81} 938

\bibitem{YJL11} Lin Y-J, Jim\'{e}nez-Garc\'{\i}a K and Spielman I B 2011
\textit{Nature} \textbf{471} 83

\bibitem{MG1205} Gong M, Qian Y-Y, Scarola V W and Zhang C-W 2012 \textit{%
Sci. Rep.} \textbf{5} 10050

\bibitem{MO90} Ogata M and Shiba H 1990 \textit{Phys. Rev. B} \textbf{41}
2326

\bibitem{CCC08} Chang C-C and Zhang S 2008 \textit{Phys. Rev. B} \textbf{78}
165101

\bibitem{MP14} Piraud M, Cai Z, McCulloch I P and Schollw\"{o}ck U 2014
\textit{Phys. Rev. A} \textbf{89} 063618

\bibitem{SP14} Peotta S, Mazza L, Vicari E, Polini M, Fazio R and Rossini D
2014 \textit{J. Stat. Mech.} P09005

\bibitem{GGBOOK} Giuliani G and Vignale G 2005 \textit{Quantum theory of the
electron liquid} (Cambridge: Cambridge University Press)

\bibitem{ZQY11} Yu Z-Q and Zhai H 2011 \textit{Phys. Rev. Lett.} \textbf{107}
195305

\bibitem{GC12} Chen G, Gong M and Zhang C-W 2012 \textit{Phys. Rev. A}
\textbf{85} 013601

\bibitem{SD07} Dutta S, Lakshmi S and Pati S K 2007 \textit{J. Phys.:
Condens. Matter} \textbf{19} 322201

\bibitem{KB07} Buchta K, Legeza \"{O}, Szirmai E and S\'{o}lyom J 2007
\textit{Phys. Rev. B} \textbf{75} 155108

\bibitem{JEH83} Hirsch J E and Scalapino D J 1983 \textit{Phys. Rev. B}
\textbf{27} 7169
\end{thebibliography}
\end{document}